\documentclass[twocol]{ametsocV6.1}

\title{The persistence of a tropical cyclone seed vortex depends on its structure\\
{\color{red}NOT PUBLISHED. Submitted for peer review}}

\authors{Kuan-Yu Lu,\aff{a}\correspondingauthor{Kuan-Yu Lu, kuanyulu.as@gmail.com}
Daniel R. Chavas,\aff{a} 
Danyang Wang \aff{a}}

\affiliation{\aff{a}{Department of Earth, Atmospheric, and Planetary Sciences, Purdue University, West Lafayette, Indiana}}

\abstract{Tropical cyclones (TC) are often generated from pre-existing ``seed" vortices. Seeds with higher persistence might have a higher chance to undergo TC genesis. What controls seed persistence remains unclear. This study proposes that planetary Rossby wave drag is a key factor that affects seed persistence. Using recently developed theory for the response of a vortex to the planetary vorticity gradient, a new parameter given by the ratio of the maximum wind speed ($V_{max}$) to the Rhines speed at the radius of maximum wind ($R_{max}$), here termed ``vortex compactness" ($C_v$), is introduced to characterize the vortex weakening by planetary Rossby wave drag. The relationship between vortex compactness and weakening rate is tested via barotropic $\beta$-plane experiments. The vortex's initial $C_v$ is varied by systematically varying their initial $V_{max}$ and $R_{max}$ in idealized wind profile models. Experiments using vorticity distributions from real-world seeds possessing are also taken from reanalysis. The weakening rate depends strongly on the vortex's initial $C_v$ across both idealized and real-world experiments, and the asymmetry introduces minor differences. Experiments doubling the size of seed vortices cause them to weaken more rapidly in line with other experiment sets. The dependence of the weakening rate on initial compactness can be predicted from a simple theory, which is more robust for more compact vortices.
Our results suggest that a seed's structure strongly modulates how long it can persist in the presence of a planetary vorticity gradient. Connections to real seeds on Earth are discussed.}

\begin{document}

\maketitle

%
%
%
\statement
This study explores the evolution of tropical cyclones (TCs) seeds, which are pre-existing weakly rotating rainstorms, in a simple setting that isolates the dynamical effects of the rotating sphere. It is not clear why some seeds can persist for a longer duration and might have a higher chance to eventually undergo genesis. We proposed that a factor called ``planetary Rossby wave drag" plays a crucial role in this process. To investigate this, we introduce a new parameter called ``compactness" to describe how the size and intensity of a seed vortex determines how quickly it will weaken due to this drag. We conducted experiments with numerical simulations and real-world TC seeds data to test our ideas.
Our findings show that the initial compactness of seeds strongly influences how quickly it weakens. We've developed a formula to predict how quickly these patterns weaken based on their compactness, which is especially accurate for more compact seeds. This research helps us understand how planetary Rossby wave drag affects the persistence of TC seed and, ultimately, how it might impact the frequency of TCs.

%
\section{Introduction}

Tropical cyclones (TCs) are one of the most devastating natural disasters on Earth. These intense rotating storms can cause significant damage to coastlines and populations in their path \citep{Rappaport_2014, Emanuel_2005}.
Given that the frequency of TCs – defined as the occurrence counts of TCs within a specified time window and geographical domain – directly controls the overall level of hazard and risk associated with TCs, comprehending the frequency and genesis processes of TCs becomes crucial for predicting and mitigating their impacts \citep{Sobel_2021}.

Theoretically, TCs can genesis spontaneously in certain rotational radiative-convective equilibrium (RCE) simulations with perturbed convective favorable environments \citep[e.g.][]{Nolan_et_al_2007, Held_and_Zhao_2008, Wing_et_al_2016, Cronin_and_Chavas_2019}.
However, observational studies have revealed that the majority of TCs originate from pre-existing precursor disturbances, such as the intertropical convergence zone (ITCZ) breakdown \citep{Cao_et_al_2013, Kieu_and_Zhang_2008} or Africa Easterly Waves (AEWs) \citep{Thorncroft_and_Hodges_2001, Russell_et_al_2017}.
These precursor disturbances are often referred as TC ``seeds". 
Recent research proposes a seeding-transition framework to interpret TC genesis variability across different climatology \citep{Hsieh_et_al_2020, Vecchi_et_al_2019}.
This framework involves the transformation of a tropical convective cluster into a seed, and the development of a seed into a genesis of TC.
The length of the seeding stage defines the persistence of seeds. 
If a seed manages to persist and encounters a favorable thermodynamic environment (low ventilation) before dissipation, it may have a higher chance of undergoing genesis and developing into a TC \citep{Hoogewind_et_al_2020, Hsieh_et_al_2020, Tang_and_Camargo_2014, Vecchi_et_al_2019}.
In previous studies \citep[e.g.][]{Hopsch_et_al_2010, Ikehata_and_Satoh_2021}, the main focus has been on the transition probability from seeds to TCs, with relatively less attention given to the duration of the seeding stage.
Moreover, past investigations \citep{Lee_et_al_2020, Hsieh_et_al_2020, Vecchi_et_al_2019, Yamada_et_al_2021, Sugi_et_al_2020} have primarily concentrated on the influence of environmental factors on seed development, absolute counts, and transition probability.
However, there has been limited inquiry into how the structure and dynamics of the seed itself affect its persistence.

Theoretically, seed persistence can affect the frequency of TCs in two ways.
First, a seed that persists for a longer duration has a higher probability of encountering an environment favorable for the genesis of TCs, assuming the climatology of the ventilation index remains unchanged.
Secondly, increasing the overall persistence of seeds contributes to higher seed counts at any given point during the TC season.
Consequently, even if the generation rate remains constant, the annual frequency of TCs may still increase due to the augmented number of seeds.
Therefore, investigating the dynamics of seed persistence leads to a more comprehensive understanding of TC frequency.

\begin{figure*}[ht]
 \centerline{\includegraphics[width=40pc]{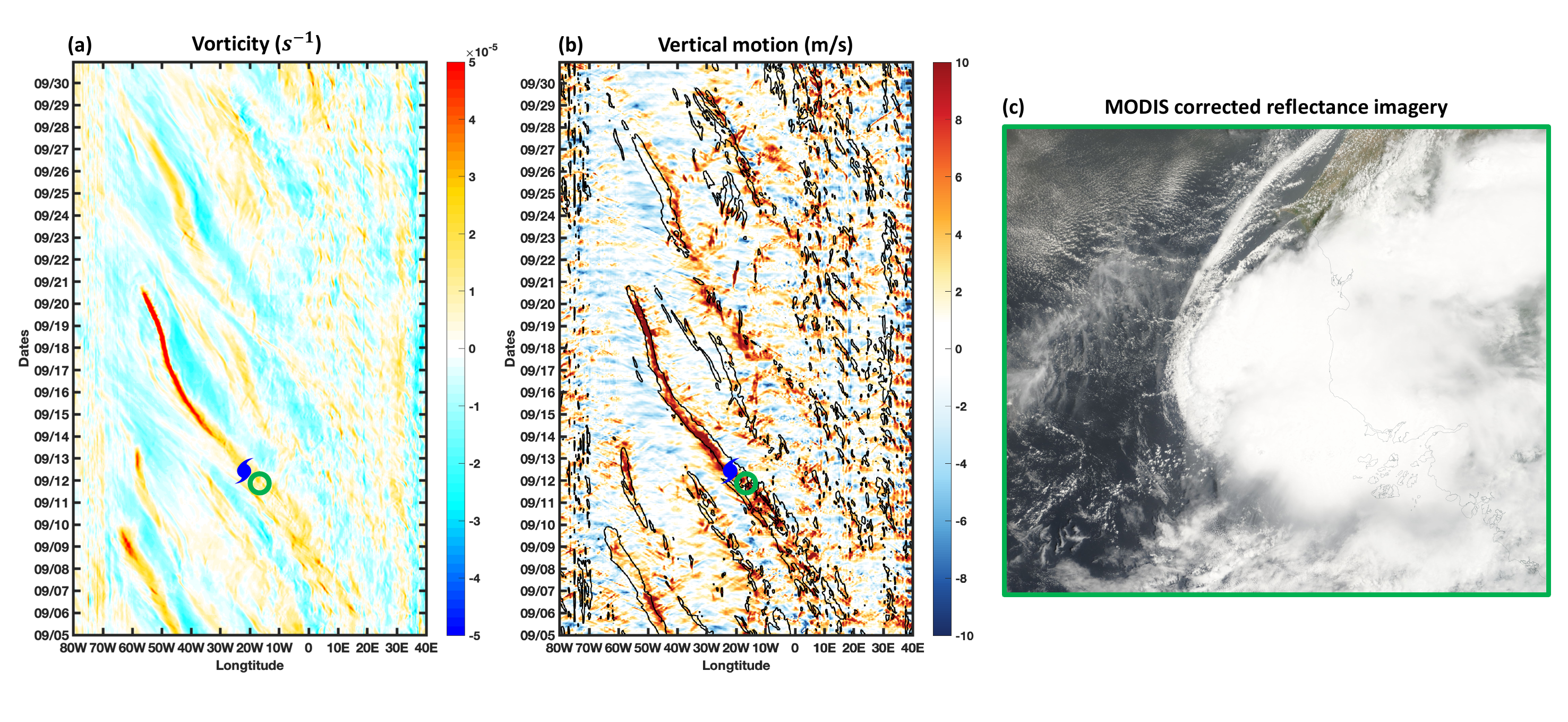}}
  \caption{ An example of African easterly waves that generate TC seeds.
  (a) The Hovmöller diagram illustrates the evolution of ERA5 $850$-hPa relative vorticity averaged between $5^\circ$ and $15^\circ$N.
  The blue cyclone symbol represents the genesis of Hurricane Helene (2006).
  (b) Same as (a), but for the average vertical motion between $1000$ and $500$-hPa.
  Black contours indicate averaged $850$-hPa relative vorticity greater than $10^{-5} s^{-1}$.
  (c) A MODIS corrected reflectance satellite image of Hurricane Helene's seed.
  The satellite image was downloaded from NASA EOSDIS Worldview.
  The location and time of this snapshot are indicated by the green circle in both (a) and (b).}
  \label{fig:Example_Seed}
\end{figure*}

As mentioned previously, TC seeds typically originate from a variety of larger-scale dynamical phenomena. 
For instance, in the North Atlantic basin, African Easterly Waves (AEWs) serve as the primary source of TC seeds \citep{Thorncroft_and_Hodges_2001, Russell_et_al_2017}.
Fig.\ref{fig:Example_Seed} illustrates the AEW seed of Hurricane Helene (2006) along with other nearby AEWs, the latter of which are all potential (but non-developing) TC seeds that persist for varying durations.
The figure suggests that even seeds originating from the same larger-scale dynamical phenomenon around a similar time can exhibit different persistence, and there is limited discussion of this variability.

The meridional variance in the Coriolis parameter on Earth is commonly denoted as $\beta$.
Recently, \citet[][hereafter LC22]{Lu_and_Chavas_2022} proposed that the size of a TC-like vortex on a barotropic $\beta$-plane is constrained by the vortex Rhines scale ($R_{VRS}$).
Specifically, circulations larger than $R_{VRS}$ dissipate rapidly due to the influence of planetary Rossby waves, while circulations within $R_{VRS}$ remain largely unaffected and maintain axisymmetry.
LC22's primary focus lies on examining the size of mature TC-like vortices, which typically possess the strongest circulation located well within $R_{VRS}$.
The impact of planetary Rossby waves on vortex intensity remains uncertain when the strongest circulation of a vortex is situated near or even outside $R_{VRS}$.

Based on the previous introduction, here we are proposing following research questions:
\begin{enumerate}
    \item How does planetary Rossby wave affect the intensity of an individual seed vortex?
    \item Does the structure of the vortex affect the vortex intensity response to planetary Rossby wave?
\end{enumerate}
To answer these questions, Section \ref{Theory} will begin by revisiting the Rhines effect as proposed in LC22 and subsequently establish a novel structural parameter to characterize the influence of planetary Rossby waves on vortex intensity. 
Next, Section \ref{Methods} will demonstrate the model configuration, data processing, and experimental designs employed in this study. 
The results and analyses of our experimental sets will be presented in Section \ref{Results}. 
Finally, Section \ref{Conclusions} will provide a summary of our key findings, along with a discussion of the implications arising from our results.

\section{Theoretical background}
\label{Theory}

\subsection{Planetary Rossby wave drag, Rhines number, and dynamical pouch}
The essence of a TC seed is simply a rotating convective cluster, and its development can be expressed by a low-level vorticity equation for flow above the boundary layer, away from friction \citep{Hsieh_et_al_2020, Raymond_and_Lopez_2011, Haynes_and_McIntyre_1987}:
  \begin{equation}\label{eq:Vor_E}
    \frac{d\zeta}{dt}=(-\delta)(f+\zeta)-(\beta+\partial_y\bar{\zeta})v+\varepsilon,
  \end{equation}
where $\zeta$ represents the relative vertical vorticity, $f$ denotes the Coriolis parameter, $\bar{\zeta}$ is the environmental (background mean) vorticity, $\delta$ stands for horizontal divergence, $\beta$ represents the meridional gradient of $f$, $\partial_y$ is the meridional gradient operator, $v$ is the meridional wind, and $\varepsilon$ is small-scale vorticity forcing term such as the baroclinic generation term. 
The physical meaning of the first term on the left-hand side ($LHS$) is the Lagrangian tendency (following the vortex) of relative vertical vorticity. 
The first term on the right-hand side ($RHS$) represents the stretching term of absolute vorticity. 
The second term on the $RHS$ includes the meridional advection of planetary vorticity ($-\beta v$) and environmental vorticity ($-\partial_y\bar{\zeta}v$), whose sum is often referred to as the effective $\beta$-term.

A typical TC seed is characterized by cyclonic rotation and active convection, leading to a low-level convergence.
As a consequence, the stretching term within a TC seed typically acts as a vorticity source term.
On the other hand, the effective $\beta$ (i.e., the meridional gradient of absolute vorticity), represented by $\beta+\partial_y\bar{\zeta}$ (numerator of Eq.\ref{eq:Vor_E}), provides a background meridional vorticity gradient, stimulating planetary Rossby waves and decreasing vorticity through wave drag effects.
Therefore, this $\beta$ term physically represents planetary Rossby wave drag, and it is crucial for understanding the evolution of seed intensity. It is important to note that the focus of this work is on comprehending the role of the sink term (planetary Rossby wave drag) in the seed's intensity evolution. Considering the source term will be a natural progression for future research.

Unlike the typical wave drag experienced by a sailing boat or human swimming, which dissipates linear momentum by generating gravity waves \citep{Vorontsov_and_Rumyantsev_2000, Wilson_and_Thorp_2003, Vennell_et_al_2006, Yang_et_al_2013}, planetary Rossby wave drag dissipates angular momentum by generating planetary Rossby waves, thus decelerating the circulation of a vortex. 
Furthermore, planetary Rossby wave drag is not radially homogeneous within a vortex.
To demonstrate the radial dependence of planetary Rossby wave drag within a vortex, we adopt a 2-D barotropic vorticity equation, which is the simplest model that allows for the consideration of planetary Rossby wave drag.
By setting $\delta=0$, omitting $\varepsilon$, assuming $\partial_y\bar{\zeta}=0$, and only considering horizontal components in Eq.\ref{eq:Vor_E}, we can obtain the 2-D barotropic vorticity equation:
By setting $\delta=0$, omitting $\varepsilon$, assuming $\partial_y\bar{\zeta}=0$, and only considering horizontal components in Eq.\ref{eq:Vor_E}, we obtain the 2-D barotropic vorticity equation:
  \begin{equation}\label{eq:Barotropic_Vor}
    \frac{\partial \zeta}{\partial t} = -\overrightarrow{\textbf{u}} \cdot \bigtriangledown \zeta -\beta v,
  \end{equation}
where $\overrightarrow{\textbf{u}}$ represents the horizontal wind field. 
In Eq.\ref{eq:Barotropic_Vor}, $LHS$ is the vorticity Eulerian tendency term. 
The first term on $RHS$ is the non-linear vorticity advection term, and the second term represents the $\beta$-term.

When $\beta$-term dominates the vorticity tendency in Eq.\ref{eq:Barotropic_Vor}, it stimulates planetary Rossby waves. To express the relationship between $\beta$-term and the non-linear advection term within a vortex, LC22 defines the ratio between these two terms as the Rhines number, $Rh$. By converting the coordinate into a cylindrical form, applying scale analysis and $\beta$-plane approximation to Eq.\ref{eq:Barotropic_Vor}, $Rh$ can be written as follows:
  \begin{equation}\label{eq:Rhines_number}
    \frac{\overrightarrow{\textbf{u}} \cdot \bigtriangledown \zeta}{\beta v} \equiv Rh=\frac{U_t}{2 \pi \beta r^2}.
  \end{equation}
where $U_t$ is the tangential circulation speed of a vortex, and $r$ is the radius of the circulation. A detailed derivation is described in LC22. When a circulation has a faster speed (stronger $U_t$) or a smaller size (smaller $r$), it results in a larger $Rh$ value ($Rh \gg 1$). This indicates that the circulation is not dominated by the $\beta$ term and can circulate without a significant amount of planetary Rossby wave drag on it. On the other hand, if circulation is slower (has a smaller $U_t$) or has a larger size (larger $r$), it results in a smaller $Rh$ value ($Rh \ll 1$), indicating that the circulation is dominated by the $\beta$ term and is significantly decelerated by planetary Rossby wave drag.

\begin{figure*}[ht]
 \centerline{\includegraphics[width=40pc]{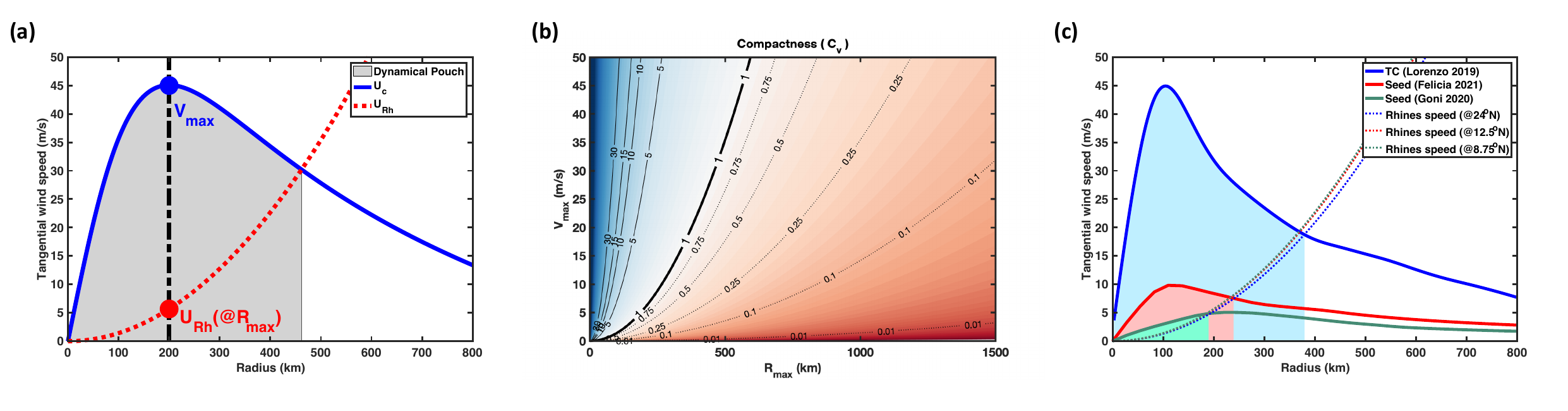}}
  \caption{
  A comprehensive visualization of the definition of $C_v$, the dynamical pouch, variations in $C_v$, and three example wind profiles with different $C_v$ values.
  (a) The conceptual diagram illustrates the definition of $C_v$ and the dynamical pouch for an example wind profile. 
  The blue curve represents the azimuthal averaged tangential wind profile of an idealized mature TC-like vortex, while the red dashed curve indicates the $U_{Rh}$ profile when $\beta=\beta(@10^{\circ}N)$.
  The gray shading area indicates the region of the dynamical pouch, which is the region of the vortex lying inside the vortex Rhines scale (depicted by the black solid vertical line).
  (b) The conceptual diagram illustrates $C_v$ corresponding to different $V_{max}$ and $R_{max}$.
  Warmer colors indicate larger $C_v$, while colder colors represent smaller $C_v$.
  The black curves are the contours of $C_v$, with solid curves indicating $C_v>1$, and dashed curves indicating $C_v<1$.
  (c) The tangential wind profiles of an example mature TC (blue), an example compact seed (red), and an example incompact seed (green).
  The mature TC structure is obtained from the ERA5 $850$-hPa relative vorticity of Hurricane Lorenzo (2019) at its lifetime peak intensity.
  The compact seed corresponds to the seed of Hurricane Felicia (2021), and the incompact seed corresponds to the seed of Cyclone Goni (2020). 
  Both seeds are obtained at their lifetime peak intensity.
  Solid curves represent their tangential wind profiles, and dashed curves represent their corresponding $U_{Rh}$ profiles. 
  The dynamical pouch region of each vortex is indicated by different colored shading.}
  \label{fig:theory}
\end{figure*}

As demonstrated in LC22, we may define the Rhines speed, $U_{Rh}$, by setting $Rh=1$ in Eq.\ref{eq:Rhines_number} and solving for $U_t$, resulting in:
  \begin{equation}\label{eq:RhinesSpeed}
    {U}_{Rh}(r)\equiv 2 \pi \beta {r}^2.
  \end{equation}
The $U_{Rh}$ profile (red dashed curve in Fig.\ref{fig:theory}a) on a $\beta$-plane is fixed because it depends solely on $r$ and $\beta$. 
For a circulation with a known size, $U_{Rh}$ represents a transition speed that determines the significance of planetary Rossby wave drag. 
That is, planetary Rossby wave drag is strong when $U_t<U_{Rh}$ and is weak when $U_t>U_{Rh}$. 
In LC22, the radius of the intersection between $U_{Rh}$ and $U_t$ profiles is defined as the vortex Rhines scale ($R_{VRS}$), depicted as the black solid vertical line in Fig.\ref{fig:theory}a. 
Analogous to the concept of a ``thermodynamic pouch" proposed by \cite{Dunkerton_et_al_2009}, we refer to the region within $R_{VRS}$ as the ``dynamical pouch" (highlighted in the shaded region in Fig.\ref{fig:theory}a).
Within this dynamical pouch, the circulation speed is faster than $U_{Rh}$, and it is expected to be ``protected" from the influence of planetary Rossby wave drag. 
However, it is important to note that while we can mathematically define a clear boundary for the dynamical pouch, it should be viewed physically as a transition zone between the inner pouch region and the outer wave drag region.

LC22 investigated this theory within the framework of strong mature TC-like vortices where their azimuthally averaged maximum wind speed ($V_{max}$) is significantly larger than $U_{Rh}$ at the radius of $V_{max}$ ($R_{max}$), resulting in planetary Rossby wave drag primarily affecting the outer circulation and constraining storm size while maintaining a steady $V_{max}$. In this study, our focus is on weaker seed vortices that do not meet this criterion, potentially leading to more planetary Rossby wave drag on $V_{max}$, and hence weakening of the seed vortex, due to the influence of planetary Rossby wave drag.

\subsection{Structural compactness}

A seed vortex typically exhibits a lower $V_{max}$ and/or a larger $R_{max}$. When these conditions are met, the seed's $V_{max}$ is not significantly larger, sometimes even smaller than $U_{Rh}$ at $R_{max}$. As a consequence, the vortex's intensity is no longer distinctly protected by the dynamical pouch, and the impact of planetary Rossby wave drag on the seed's $V_{max}$ becomes a significant consideration. To precisely quantify the effect of planetary Rossby wave drag on the vortex's intensity, we introduce the concept of vortex structural compactness, denoted as $C_v$. This structural parameter characterizes the ratio between $V_{max}$ and $U_{Rh}$ at $R_{max}$:
  \begin{equation}\label{eq:Compactness}
    C_v\equiv Rh(@R_{max})=\frac{V_{max}}{U_{Rh}(@R_{max})}=\frac{V_{max}}{2\pi\beta R_{max}^2}.
  \end{equation}
Fig.\ref{fig:theory}a provides a visual representation of the definition of $C_v$. Importantly, $C_v$ varies with both $V_{max}$ and $R_{max}$. As illustrated in Fig.\ref{fig:theory}b, a vortex characterized by a stronger $V_{max}$ and/or a smaller $R_{max}$ has a compact structure ($C_v \gg 1$). On the other hand, a vortex with a weaker $V_{max}$ and/or a larger $R_{max}$ has an incompact structure ($C_v \ll 1$). Moreover, when the vortex possesses multiple local maxima of $U_t$ values, $C_v$ solely accounts for the maximum that yields the highest $C_v$. It is also essential to note that the transition region near the $C_v=1$ contour in Fig.\ref{fig:theory}b emphasizes the absence of a distinct boundary between compact and incompact structures.

Figure \ref{fig:theory}c shows the tangential wind profiles of three example vortices obtained from ERA5 reanalysis: Hurricane Lorenzo (2019) as an example mature TC, the seed of Hurricane Felicia (2021) as an example compact seed, and the seed of Typhoon Goni (2020) as an example incompact seed.
For the mature Hurricane Lorenzo (blue curve in Fig. \ref{fig:theory}c), its $V_{max}$ is significantly greater than $U_{Rh}$ at $R_{max}$, resulting in a highly compact structure ($C_v = 28.927$). 
Consequently, the intensity of Hurricane Lorenzo is not expected to be significantly affected by planetary Rossby wave drag. 
Note that the inner core of Hurricane Lorenzo is certainly under-resolved in ERA5, but this does not alter our interpretation.

In comparison, the seed of Hurricane Felicia (red curve in Fig. \ref{fig:theory}c) exhibits a much weaker $V_{max}$ and a similar $R_{max}$, resulting in a relatively smaller $C_v = 5.7846$, although it still exceeds $1$. 
Consequently, the seed vortex of Hurricane Felicia is classified as a compact seed ($C_v > 1$), but it is expected to experience a greater weakening due to planetary Rossby wave drag compared to Hurricane Lorenzo.
On the other hand, the seed of Typhoon Goni displays the weakest $V_{max}$ and the largest $R_{max}$ among these three profiles, resulting in the smallest $C_v = 0.7361$, which falls even below $1$ (green curve in Fig. \ref{fig:theory}c). 
Therefore, the seed of Typhoon Goni is anticipated to be subjected to the most substantial influence of planetary Rossby wave drag on its intensity when compared to the other two compact vortices.

We would like to emphasize that our theory does not provide an explanation for the specific factors influencing a vortex's value of $C_v$ (i.e., how the vortex and its specific structure formed in the first place).
Its sole purpose is to describe the impact of planetary Rossby wave drag on the vortex's intensity.
Furthermore, the derivation of $C_v$ is based on the barotropic $\beta$-plane assumption and does not consider the vorticity source term due to stretching or other factors, as our focus lies in understanding the role of the sink term (planetary Rossby wave drag) acting alone.
Therefore, while $C_v$ directly represents the influence of planetary Rossby wave drag on $V_{max}$, it cannot solely determine the intensity tendencies of vortices in real-world scenarios.

\subsection{Hypothesis}
According to our theory, $C_v$ governs planetary Rossby wave drag on $V_{max}$, thereby directly influencing the weakening of a vortex due to planetary Rossby wave drag.
To investigate the impact of $C_v$ on planetary Rossby wave drag on vortex intensity, as well as its effect on the persistence of TC seed vortex, we propose the following hypothesis: A more compact vortex (a higher $C_v$) will experience a slower weakening on a barotropic $\beta$-plane.
We will test this hypothesis by conducting experiments using a barotropic $\beta$-plane model.
The model will be initialized with both a idealized axisymmetric vortex model and real-world seed vorticity structures obtained from reanalysis data.

\section{Methods}
\label{Methods}

\subsection{Barotropic model}
To examine the isolated impact of planetary Rossby wave drag on the vortex intensity, we employ a barotropic $\beta$-plane model developed by James Penn and Geoffrey K. Vallis (available at: https://empslocal.ex.ac.uk/people/staff/gv219/codes/barovort.html).
This model exclusively incorporates planetary Rossby waves as a vorticity sinking process, with no inclusion of any source term.
The model utilizes a pseudospectral method to solve Eq.\ref{eq:Barotropic_Vor} in 2-D space.
The pseudospectral method employs analytic derivatives to calculate horizontal winds, which are then used to evaluate nonlinear advection terms in Eq.\ref{eq:Barotropic_Vor}.
The model has $500$ grid points in both the $x$ and $y$ directions, with a grid spacing of $20$ km.
The initial time-step is set at $60$ seconds, and the model also employs an adaptive time-step to meet the Courant–Friedrichs–Lewy (CFL) condition.
There is no external forcing applied in any of our experiments.
To ensure numerical stability, the model employs a dissipation process known as the high wave-number Smith filter \citep{Smith_et_al_2002}.
This filter damps any structures with wave-numbers exceeding $30$.

It is important to note that the barotropic model employed in this study is two-dimensional and non-divergent.
Consequently, it does not account for the radial flow and vertical motions typically observed in a real convective TC seed.
As a result, the barotropic model neglects the azimuthal-mean radial mass transport while still allowing for eddies to radially transport momentum.
By doing so, the model can effectively simulate the vortex response to planetary Rossby wave drag, while minimizing interactions across different radii within the vortex.
Accounting for the role of inflow and the vorticity source term (stretching) is a critical next step in future work.
Lastly, we utilize the centroid of vorticity as a tracking metric to determine the vortex center over time.
By employing this center, we calculate the azimuthally averaged radial profiles of tangential wind for every experiment.

\begin{table*}[h]
\caption{Selected TCs' information from IBTrACS.}\label{t1}
\begin{center}
\begin{tabular}{cccccc}
Basin             & TC name  & ISO Time            & Initial Latitude & Initial Longitude &  \\ \cline{1-5}
Northern Indian   & Amphan   & 2020-05-15 06:00:00 & 9.5 N            & 87.5 E            &  \\
Eastern Pacific   & Douglas  & 2020-07-20 00:00:00 & 14.7 N           & 118.8 W           &  \\
Northern Atlantic & Eta      & 2020-10-31 18:00:00 & 14.9 N           & 72.4 W            &  \\
Northern Indian   & Gati     & 2020-11-20 12:00:00 & 11.2 N           & 62.6 E            &  \\
Western Pacific   & Goni     & 2020-10-25 06:00:00 & 10.6 N           & 143.9 E           &  \\
Western Pacific   & Haishen  & 2020-08-30 12:00:00 & 24.9 N           & 145.5 E           &  \\
Northern Atlantic & Laura    & 2020-08-20 00:00:00 & 14.4 N           & 47.3 W            &  \\
Eastern Pacific   & Marie    & 2020-09-27 12:00:00 & 12.4 N           & 100.4 W           &  \\
Western Pacific   & Maysak   & 2020-08-26 00:00:00 & 12.8 N           & 133.9 E           &  \\
Western Pacific   & Vamco    & 2020-11-08 00:00:00 & 8 N              & 134.9 E           &  \\
Western Pacific   & Chanthu  & 2021-09-05 06:00:00 & 12.4 N           & 140.5 E           &  \\
Eastern Pacific   & Felicia  & 2021-07-14 00:00:00 & 13.5 N           & 110.6 W           &  \\
Northern Atlantic & Grace    & 2021-08-13 06:00:00 & 15 N             & 46.7 W            &  \\
Northern Atlantic & Ida      & 2021-08-26 12:00:00 & 16.5 N           & 78.9 W            &  \\
Eastern Pacific   & Linda    & 2021-08-09 12:00:00 & 12.1 N           & 98.2 W            &  \\
Western Pacific   & Mindulle & 2021-09-22 06:00:00 & 10.8 N           & 150 E             &  \\
Western Pacific   & Nyatoh   & 2021-11-28 12:00:00 & 11.2 N           & 146.2 E           &  \\
Western Pacific   & Rai      & 2021-12-11 12:00:00 & 5.3 N            & 145 E             &  \\
Northern Atlantic & Sam      & 2021-09-22 18:00:00 & 10 N             & 33.1 W            &  \\
Western Pacific   & Surigae  & 2021-04-11 12:00:00 & 5.5 N            & 143.8 E           &  \\ \cline{1-5}
\end{tabular}
\end{center}
\end{table*}

\subsection{Idealized wind field models}
To achieve varying levels of structural compactness while maintaining physical consistency in the generated vortices, we employ an idealized wind field model that allows us to modify the $V_{max}$ and $R_{max}$ of an axisymmetric vortex.
We test two different idealized wind field models for the radial profile of the vortex azimuthal wind. 
Firstly, we use the C15 model \citep{Chavas_et_al_2015} for the complete radial profile of the TC low-level tangential wind field to initialize the barotropic model.
The C15 model allows us to define the wind profile using a limited set of storm and environmental parameters.
The storm parameters include $V_{max}$, $R_{max}$, and the Coriolis parameter ($f$).
The environmental parameters encompass the radiative-subsidence rate ($w_{cool}$), the surface drag coefficient ($C_d$) for the outer region, and the ratio of surface coefficients of enthalpy and drag ($C_k/C_d$).
In this study, we fixed $f=f(@10^\circ)=2.5325 \times 10^{-5}$ $s^{-1}$, $w_{cool}=0.002$ $m/s$, $C_d=0.0015$, $C_k/C_d = 1$, $C_{dvary}=0$, $C_kC_{dvary}=0$, ${eye}_{adj}=0$, and $\alpha_{eye}=0.15$.

By adjusting the values of $R_{max}$ and $V_{max}$ within the C15 model, we can generate tangential wind profiles with varying degrees of $C_v$.
While we specifically employ a TC wind profile model in this study, it is worth mentioning that the experimental approach could theoretically be applied to any wind profile, as long as there is no significant symmetric instability.

A simplified wind profile model can effectively capture the dynamic characteristics of an actual vortex.
As demonstrated in Fig.8 of \cite{Klotzbach_et_al_2022}, a modified Rankine vortex model successfully represents the observed TC wind field between $R_{max}$ and the $34$-kt wind radius.
Hence, we have conducted a modified Rankine vortex model, which takes the following form:
\begin{equation}\label{eq:Rankine}
        U_t(r) = \left\{ \begin{array}{rcl}
        V_{max}(\frac{r}{R_{max}})(\frac{R_{max}}{r})^\lambda & \mbox{when} & r<R_{max} \\ 
        V_{max}(\frac{r-R_0}{R_{max}-R_0})(\frac{R_{max}}{r})^\chi & \mbox{when} & r\geq R_{max}
    \end{array},\right.
\end{equation}
where $\lambda$ sets the radial increasing rate of tangential winds within $R_{max}$, and $\chi$ represents the radial decreasing rate of tangential winds outside $R_{max}$.
Note that this modified Rankine model generates tangential wind profiles with different $V_{max}$ and $R_{max}$ while holding $R_0$ fixed. In this study, we set $R_0=1500$ km, $\lambda=0.2$, and $\chi=0.5$ in every Rankine vortices.

\subsection{TC seeds from reanalysis}
\begin{figure*}[ht]
 \centerline{\includegraphics[width=35pc]{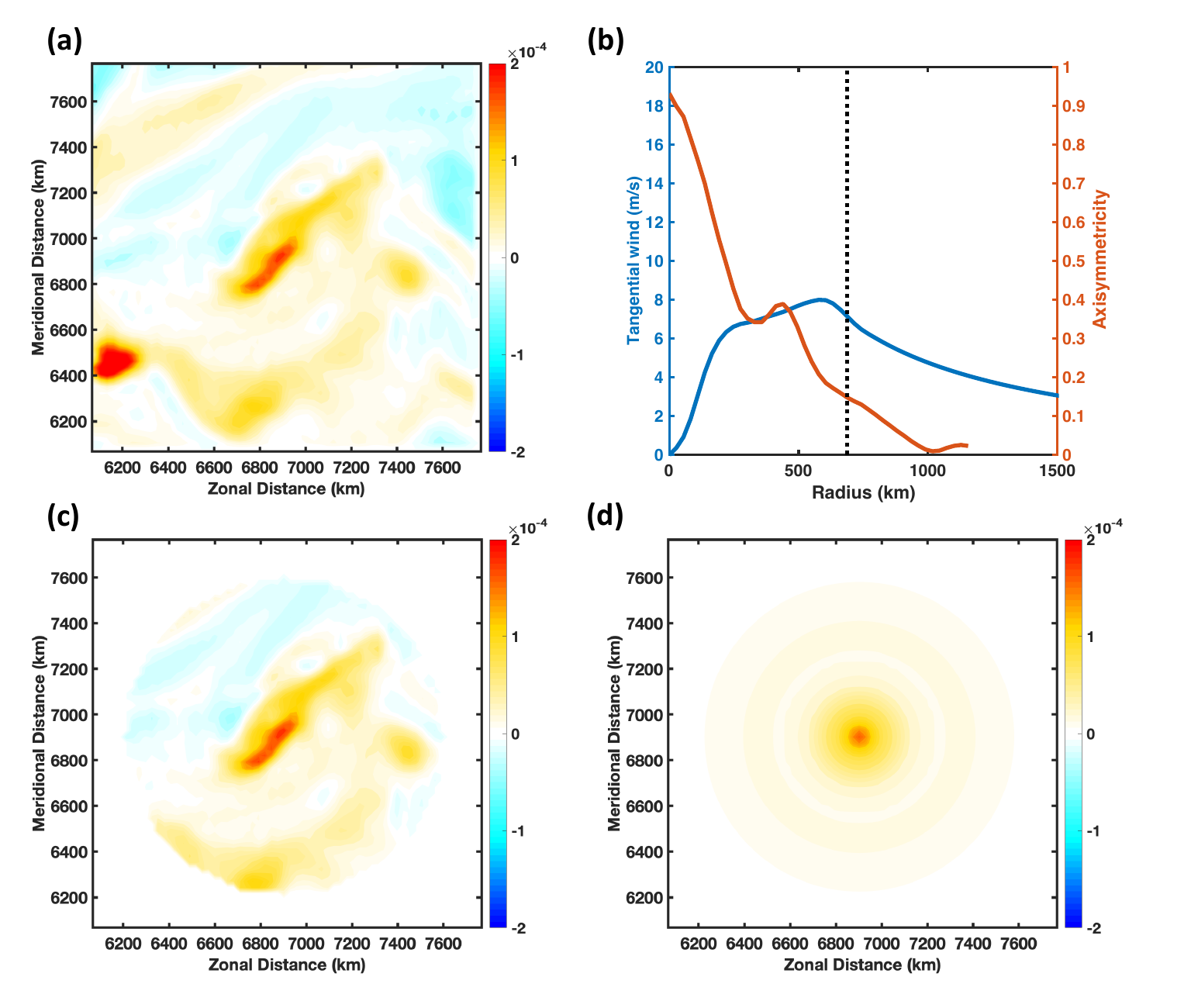}}
  \caption{
  A demonstration of the data processing for an example seed vortex (Cyclone Gati 2020) obtained from ERA5.
  (a) The $850$-hPa relative vorticity field of the seed directly acquired from ERA5.
  (b) The azimuthal averaged tangential wind profile and the axisymmetricity profile of the seed. The black dashed line indicates the radius where axisymmetricity is $0.15$.
  (c) The seed vortex after applying the axisymmetricity filter.
  (d) The symmetric vortex used for initializing $SYMSEED$, which is generated by azimuthally averaging the filtered vortex shown in (c).}
  \label{fig:ERA5_Seed}
\end{figure*}

The dynamical structure of selected TCs' seeds is obtained from the European Centre for Medium-range Weather Forecasts (ECMWF) fifth-generation reanalysis data, ERA5.
Table \ref{t1} lists $20$ selected TC cases, all of which are Category $3$, $4$, and $5$ storms sampled in 2020 and 2021 across all basins.
To begin, we determine the ending time and location of each seed track using the corresponding TC's first best track data from the International Best Track Archive for Climate Stewardship, IBTrACS \citep{Knapp_et_al_2010}.
Next, we backtrack the vorticity center of the TC seed for $2$ days by identifying the local maximum of relative vorticity at $850$hPa in ERA5.
A TC seed track is defined when detected vorticity maxima are less than $1.25^\circ$ apart.
To exclude extra-tropical storms, we disregard all tracks located beyond $30^\circ$ latitude poleward. 
The 2-D dynamical structure of the TC seed is defined as the $850$hPa relative vorticity within a range of $7.5^\circ$ from the seed center in both the zonal and meridional directions. 
Additionally, we only consider the seed's maximum intensity structure to initialize the barotropic model. Fig.\ref{fig:ERA5_Seed}a demonstrates an example of the detected seed vorticity structure from ERA5.

To exclude adjacent systems, we apply an axisymmetricity filter based on the method proposed by \citet{Shimada_et_al_2017}. 
This filter quantifies the axisymmetric nature of the seed vorticity field across radii, yielding a parameter denoted as $\gamma$. 
The value of $\gamma$ ranges from $0$ to $1$, where a higher value indicates a greater degree of axisymmetry. 
In the case of a seed vortex, the $\gamma$ value typically exhibits a monotonically decreasing with increasing radius, as depicted by the brown curve in Fig.\ref{fig:ERA5_Seed}b.
Consequently, we eliminate any vorticity structures beyond the radius where $\gamma=0.15$ (Fig.\ref{fig:ERA5_Seed}c). 
Note that to further enhance the clarity of the $\gamma$ profile, we additionally apply a moving average filter with a width of $10$ data points. 
Fig.\ref{fig:ERA5_Seed}c demonstrates the effectiveness of the axisymmetricity filter in successfully excluding adjacent systems while still preserving the natural asymmetry of the seed.
The filtered asymmetric vortex can serve as a direct initialization for the barotropic model. 
Additionally, we can compute the azimuthally averaged vorticity structure (shown in Fig. \ref{fig:ERA5_Seed}d) to initialize the symmetric experimental set, which will be elaborated upon in a subsequent section.

\subsection{Experiment design}
\subsubsection{$VARYRVMAX$: Varying vortex's $V_{max}$ and $R_{max}$ with the C15 model}
We begin by conducting an experimental set using the C15 model and the barotropic model to test our hypothesis. 
In the C15 model, we systematically vary $V_{max}$ from $10$ m/s to $50$ m/s and $R_{max}$ from $100$ km to $600$ km, resulting in $30$ TC-like vortices with different $C_v$ values. 
The initial profiles of these vortices are depicted in Fig.\ref{fig:Exp_Ini_Results}a, and a vortex with a larger initial $C_v$ value ($C_{v,0}$) is represented with a darker color.
Subsequently, we convert each tangential wind profile into an axisymmetric vorticity field to initialize the barotropic model.
The $\beta$ value in the barotropic model remains fixed at $\beta(@10^\circ N) = 2.2547 \times 10^{-11}$ $m^{-1}s^{-1}$ across all members.
Each vortex is placed at the center of the domain and simulated separately for a one-day spin-up period.
Afterward, $\beta$ is instantaneously turned on to the constant value for the subsequent $10$ days.
We then track the centroid of the vorticity in the entire domain to calculate the azimuthally averaged vortex structure. 
This set of experiments is labeled as $VARYRVMAX$.

\subsubsection{$RANKINE$: Varying vortex's $V_{max}$ and $R_{max}$ with the modified Rankine vortex model}
Next, we conduct an experiment set similar to $VARYRVMAX$, but instead, we utilize the modified Rankine vortex model to generate wind profiles.
Similar to $VARYRVMAX$, we systematically vary both $V_{max}$ and $R_{max}$ in the modified Rankine vortex model to create $30$ vortices with different $C_{v,0}$, while keeping their $R_0$ constant.
This set of experiments is labeled as $RANKINE$ and serves to investigate whether the findings from $VARYRVMAX$ remain consistent when initializing the barotropic model with a different idealized wind profile model. 
Fig.\ref{fig:Exp_Ini_Results}b illustrates the initial profiles of all members in $RANKINE$.

\subsubsection{$ASYMSEED$ and $SYMSEED$: Seed vortices from ERA5}
Instead of relying solely on idealized wind profile models to specify the vortex structure, we conducted two experiment sets by initializing the barotropic model with realistic seed vortices obtained from ERA5 reanalysis data.
The first experiment set, denoted as $ASYMSEED$, utilizes the asymmetric seed vortex directly extracted from ERA5. 
Fig.\ref{fig:Exp_Ini_Results}c displays the azimuthally averaged initial wind profiles of all members in $ASYMSEED$.
In the second experiment set, referred to as $SYMSEED$, the vorticity initial conditions of $ASYMSEED$ are azimuthally averaged (as shown in Fig.\ref{fig:ERA5_Seed}d), and the barotropic model is initialized by the symmetric vortex with the same configurations as $ASYMSEED$. 
Note that the azimuthal averaged initial wind profiles of $SYMSEED$ (not shown) are identical to those in $ASYMSEED$ (Fig.\ref{fig:Exp_Ini_Results}c).

In both $ASYMSEED$ and $SYMSEED$, each experiment adjusts the $\beta$ value in the barotropic model according to the latitude of the seed center, while maintaining it constant throughout the entire domain and the whole $10$-day simulation. 
Moreover, there is no spin-up period in either $ASYMSEED$ or $SYMSEED$.
The objective of these experiment sets is to test our hypothesis using realistic seed vortices. 
By comparing the results between $SYMSEED$ and $ASYMSEED$, we can also investigate the effect of the asymmetry feature of realistic seeds on our hypothesis.

\subsubsection{$LARGESEED$: Re-scaling symmetric seed vortices from $SYMSEED$}
\begin{figure*}[ht]
 \centerline{\includegraphics[width=40pc]{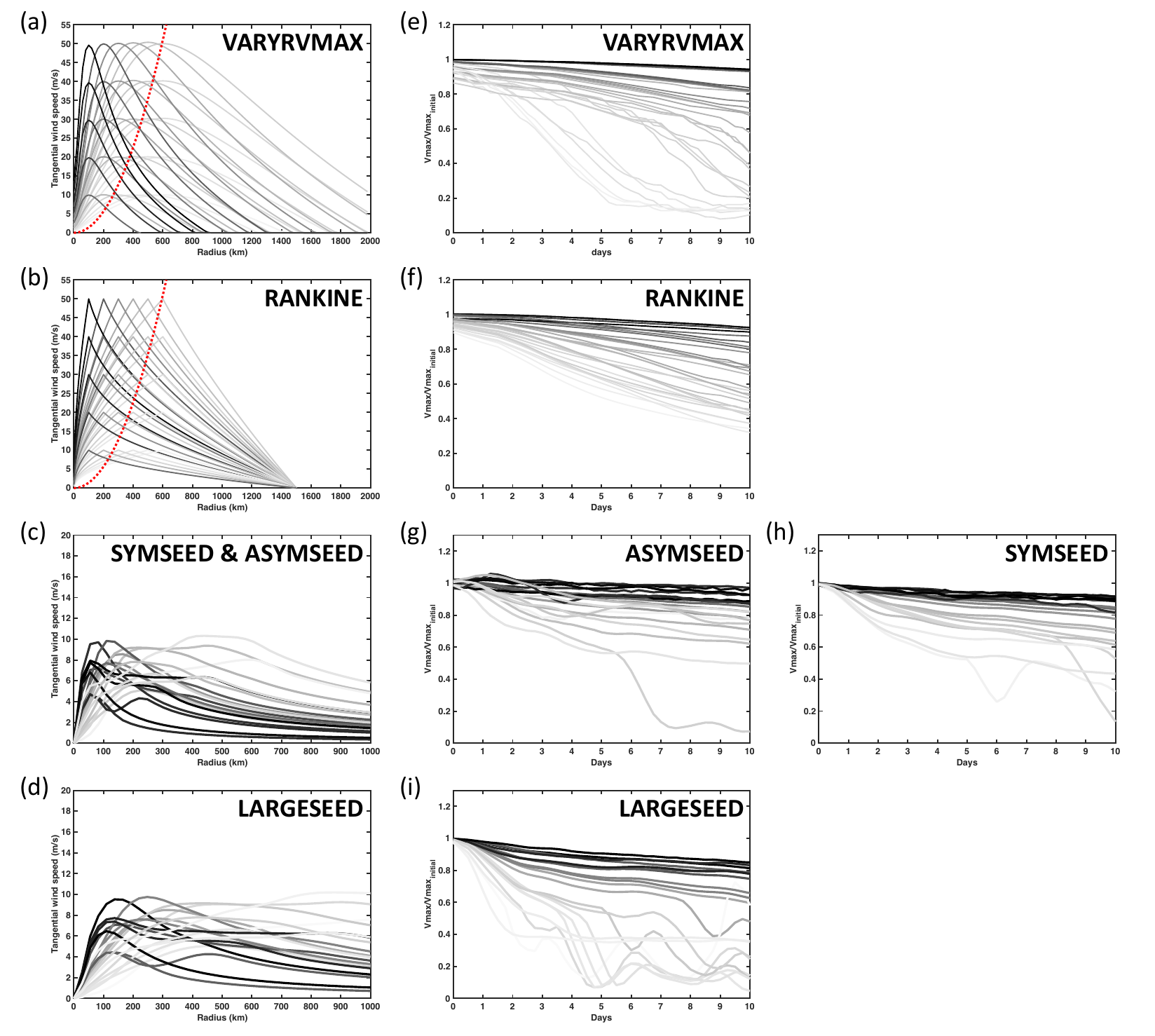}}
  \caption{
  The experiment designs and results of all our experiment sets.
  (a)-(d) The initial azimuthal averaged tangential wind profiles of all members in $VARYRVMAX$, $RANKINE$, $ASYMSEED$, and $LARGESEED$, respectively. Note that $SYMSEED$ has an identical initial wind profile as $ASYMSEED$; therefore, it is not shown.
  (e)-(i) The normalized $V_{max}$ evolution of all members in $VARYRVMAX$, $RANKINE$, $ASYMSEED$, $SYMSEED$, and $LARGESEED$, respectively. 
  The gray scale indicates the $C_{v,0}$ of each member in each experiment set, with darker colors representing relatively higher $C_{v,0}$. 
  The red dashed curves in (a) and (b) indicate $U_{Rh}$ profiles on the corresponding $\beta$-plane.
  Note that since all members in $SYMSEED$, $ASYMSEED$, and $LARGESEED$ are simulated on different $\beta$-planes, their corresponding $U_{Rh}$ profiles are not shown in (c) and (d).}
  \label{fig:Exp_Ini_Results}
\end{figure*}

Seed vortices from ERA5 naturally exhibit variability in $C_{v,0}$, but it is possible to manually decrease the vortex $C_{v,0}$ by radially enlarging the entire wind profile of a symmetrized seed vortex. 
Here, we utilize all axisymmetric members from $SYMSEED$ and enlarge their entire structure by a factor of $2$. 
This particular experiment set is referred to as $LARGESEED$, and the initial profiles are demonstrated in Fig.\ref{fig:Exp_Ini_Results}d. 
The focus of $LARGESEED$ is to test the hypothesis by reducing the $C_{v,0}$ of the seed while keeping the other aspects of the vortex structure unchanged.

\section{Results}
\label{Results}
\subsection{$VARYRVMAX$ and $RANKINE$}

Since each member has a different $V_{max}$, to standardize the comparative analysis across different members, we define the normalized intensity as $\tilde{V}_{max} = V_{max}/V_{max,0}$, where $V_{max,0}$ is the initial $V_{max}$ of each member.
Fig.\ref{fig:Exp_Ini_Results}e displays the $\tilde{V}_{max}$ evolution for all members in $VARYRVMAX$ after turning on the $\beta$. 
All members immediately start weakening after the spin-up period, including the most compact vortex (i.e., highest $C_{v,0}$). 
Overall, more compact vortices (larger $C_{v,0}$) exhibit slower rates of weakening.
Some less compact vortices (smaller $C_{v,0}$) are almost entirely dissipated by planetary Rossby wave drag, with $\tilde{V}_{max}$ evolution curves fluctuating between $0$ and $0.2$.
Fig.\ref{fig:Exp_Ini_Results}f presents similar visualizations to Fig.\ref{fig:Exp_Ini_Results}e, but for the $RANKINE$ experiment set.
Similar as in $VARYRVMAX$, more compact vortices in $RANKINE$ weaken slower.
While those compact members have similar $\tilde{V}_{max}$ evolution across the two experiment sets, less compact vortices in $RANKINE$ weaken slower than those in $VARYRVMAX$.
Another difference is that there is no member in $RANKINE$ that reaches a quasi-steady state during the $10$ days of simulation.
Results from both $VARYRVMAX$ and $RANKINE$ support the hypothesis, which states that a more compact vortex will experience less planetary Rossby wave drag on $V_{max}$ and weaken slower.
This suggests that our theory is not highly sensitive to the choice of the idealized wind profile model for the initial conditions.

\subsection{$ASYMSEED$, $SYMSEED$, and $LARGESEED$}
Fig.\ref{fig:Exp_Ini_Results}g and Fig.\ref{fig:Exp_Ini_Results}h depict the evolution of $\tilde{V}_{max}$ for $ASYMSEED$ and $SYMSEED$, respectively. 
The results obtained from both $ASYMSEED$ and $SYMSEED$ reveal that more compact vortices experience slower rates of weakening, indicating that our hypothesis is not highly sensitive to the presence of axis-asymmetry in realistic seed vortex structures.

While all members of $SYMSEED$ begin weakening immediately, certain members of $ASYMSEED$ exhibit a brief period of minor intensification at the beginning of the simulation.
This can be attributed to $ASYMSEED$'s initially asymmetric members, coupled with the absence of a spin-up period. 
Consequently, the initial axisymmetrization process of certain members in $ASYMSEED$ contributes to their minor intensification.
In contrast, the immediate weakening observed in $SYMSEED$ can be explained by its initial symmetric vortex, which bypasses any potential intensification resulting from axisymmetrization.

In addition, the majority of members in both $ASYMSEED$ and $SYMSEED$ do not weaken to a $\tilde{V}_{max}$ value below $0.4$.
The few members that reach values lower than $0.4$ do so because their maximum vorticity weakens to a magnitude similar to the stimulated planetary Rossby wave vorticity anomalies, leading to the loss of track of the centroid tracking algorithm.

The results obtained from both $ASYMSEED$ and $SYMSEED$ serve as a proof of concept for how our hypothesis applies to realistic seed structures. 
They both demonstrate that the $C_{v,0}$ remains a reliable indicator of how quickly the vortex will weaken due to planetary Rossby wave drag, and the inclusion of 2-D realistic asymmetry does not significantly diminish the reliability of $C_{v,0}$.
This is of practical use since $C_{v,0}$ can be estimated from $V_{max}$, $R_{max}$, and storm central latitude alone.

Finally, Fig.\ref{fig:Exp_Ini_Results}i illustrates the temporal evolution of $\tilde{V}_{max}$ for all members within the $LARGESEED$ experiment set.
The results from $LARGESEED$ provide robust support for our hypothesis, revealing that by manually enlarging the $SYMSEED$ members, less compact vortices weaken faster.
A comparison between $LARGESEED$ (Fig.\ref{fig:Exp_Ini_Results}i) and $SYMSEED$ (Fig. \ref{fig:Exp_Ini_Results}h) demonstrates an overall accelerated weakening rate and more entirely dissipated members in the former.
It is noteworthy that during the $10$-day simulation period, most cases in $SYMSEED$ do not attain a quasi-steady state or exhibit fluctuating behavior at low $\tilde{V}_{max}$ values.
However, two cases in the $LARGESEED$ reach a steady state, maintaining a $\tilde{V}_{max}$ around $0.4$.

\subsection{The relationship between the $C_{v,0}$ and the weakening rate}
\begin{figure*}[ht]
 \centerline{\includegraphics[width=40pc]{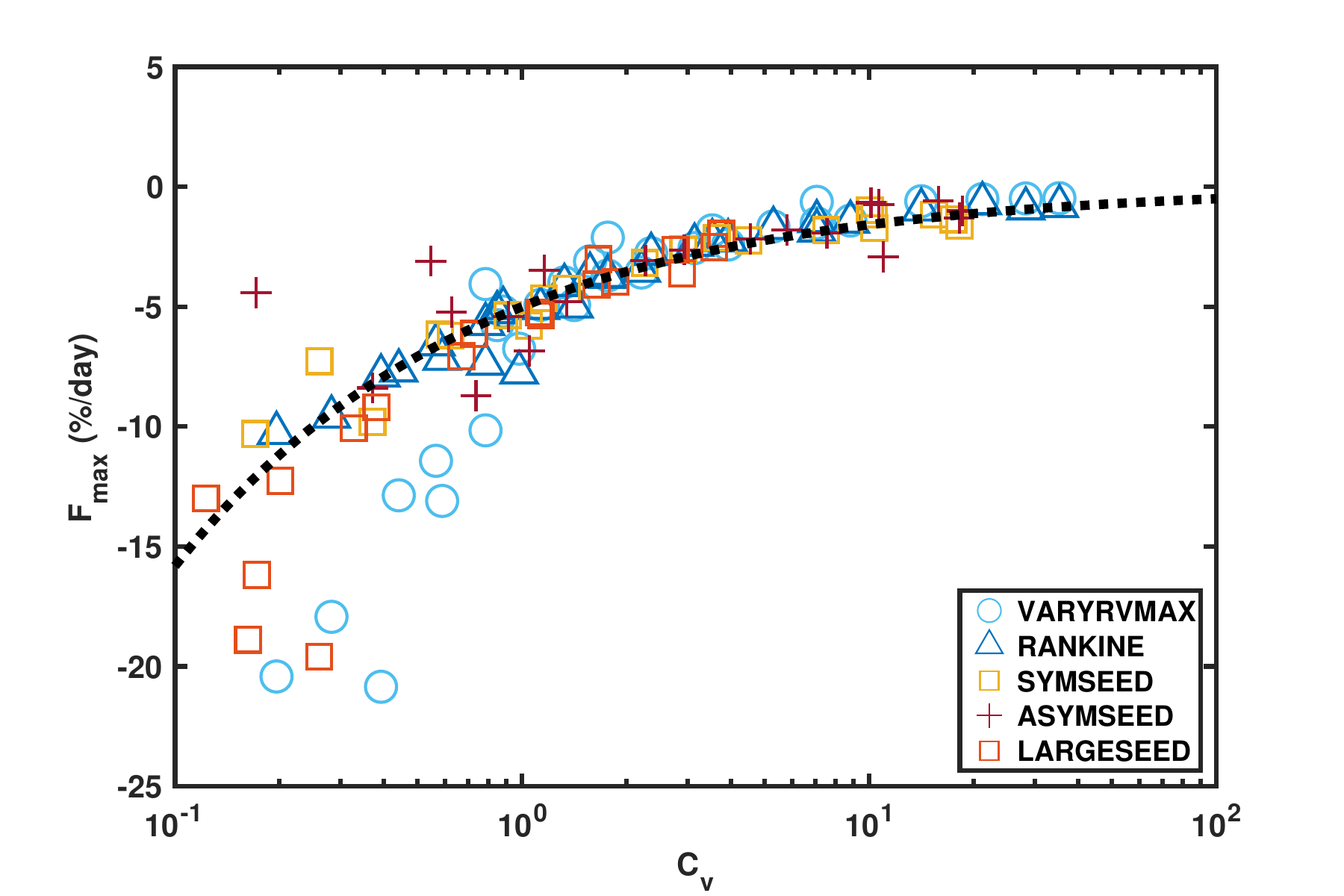}}
  \caption{
  The scatter plot illustrates the relationship between the weakening rate ($F_{max}$) and the initial structural compactness ($C_{v,0}$) of all members.
  The weakening rate $F_{max}$ is defined as the linear regression slope of the normalized $V_{max}$ evolution during the first $5$ days, expressed in units of $[\%/day]$.
  Members from the same experiment set are indicated by the same marker.
  The black dashed curve represents the fitted theoretical prediction curve.}
  \label{fig:Scatter}
\end{figure*}

As demonstrated previously, all experimental sets presented in this study support our hypothesis that vortices with higher $C_{v,0}$ values will exhibit slower weakening.
To further quantitatively validate this finding, we introduce the weakening rate, denoted as $F_{max}$, which is defined as the linear regression slope of the $\tilde{V}_{max}$ evolution during the first $5$ days of the simulation.
$F_{max}$ is expressed in units of $[\%/day]$ and represents the average daily percentage of weakening.
Fig.\ref{fig:Scatter} presents a scatter plot illustrating the relationship between the $C_{v,0}$ and $F_{max}$ values for all members across all experiment sets. 

Overall, a clear positive relationship is observed between the $C_{v,0}$ and $F_{max}$. 
Larger values of $C_{v,0}$ correspond to larger (less negative) values of $F_{max}$, indicating a slower weakening rate.
Furthermore, it seems that $F_{max}$ is inversely proportional to $-C_{v,0}^{1/2}$.
Consequently, an overlay of a prediction curve (indicated by the black dashed curve in Figure \ref{fig:Scatter}) highlights the dependence of $F_{max}$ to $C_{v,0}$. 
This prediction curve is expressed as follows:
\begin{equation}\label{eq:Fit_F_Cv}
F_{max}=-\alpha\sqrt{\frac{1}{C_{v,0}}},
\end{equation}
where $\alpha$ is a scaling factor that has the best-fitted value of $5.0$ that minimizes the $\chi^2$.
This $F_{max}$ prediction is particularly robust for the compact members across all experimental sets, as evidenced by the convergence to the prediction curve when $C_{v,0}>1$.
However, there are also distinct variations among the different experiment sets within those incompact members.
For instance, while $VARYRVMAX$ and $RANKINE$ exhibit identical distributions in $C_{v,0}$, incompact members in $VARYRVMAX$ weaken faster than the prediction curve.
On the other hand, although $SYMSEED$ and $ASYMSEED$ share identical initial azimuthal wind profiles, some incompact members in $ASYMSEED$ weakens slower than the prediction curve.
These observations suggest that the $C_{v,0}$-$F_{max}$ relationship is not significantly influenced by structural variance across the experiment sets when the vortex is compact.
However, systematic structural differences introduce more variability and decrease the robustness of the $F_{max}$ prediction for the incompact members across all experiment sets.
Furthermore, note that the results are similar when calculating $F_{max}$ over averaging periods spanning from $1$ to $8$ days.

\begin{figure*}[ht]
 \centerline{\includegraphics[width=40pc]{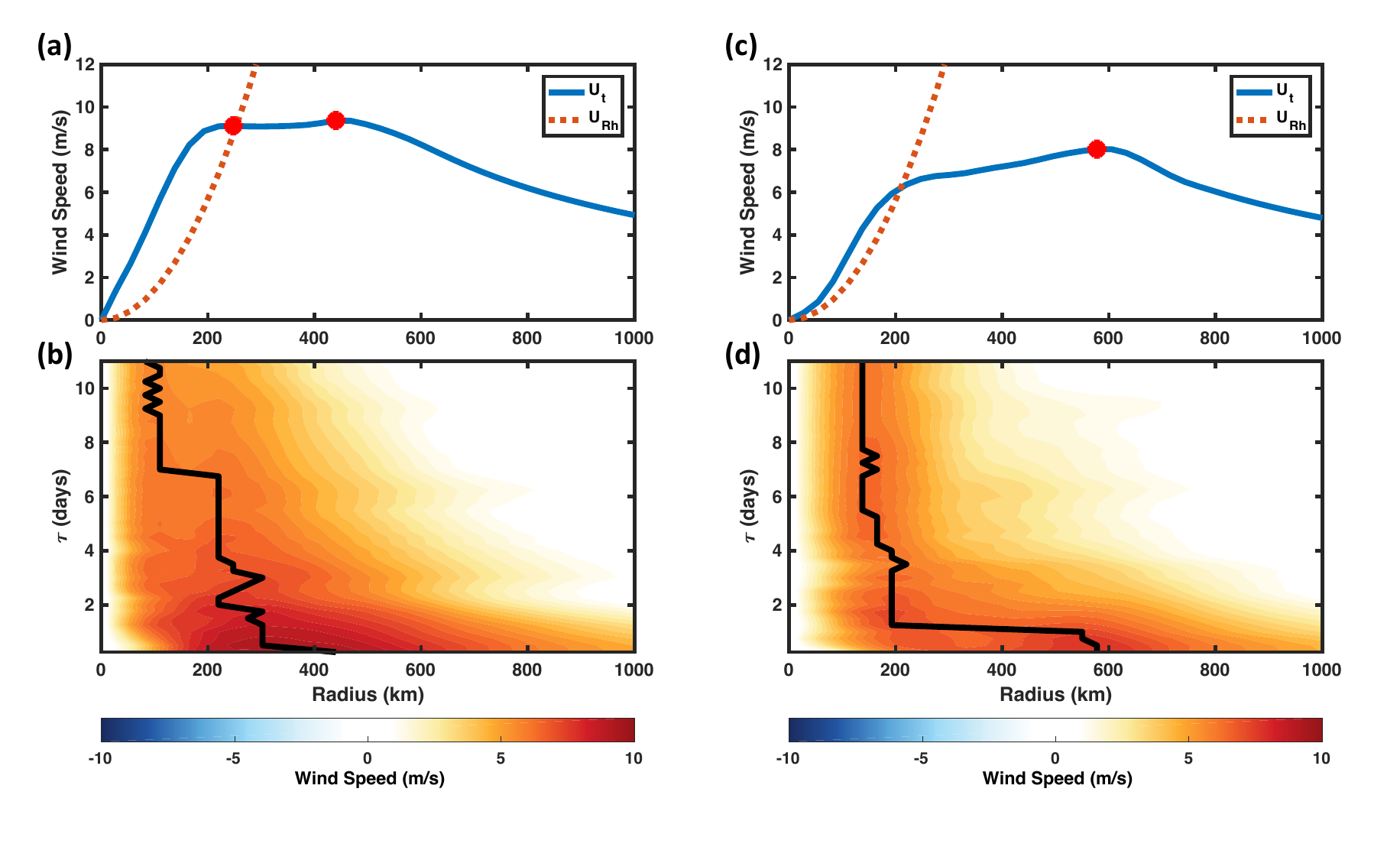}}
  \caption{
  A demonstration of two cases from $ASYMSEED$ that both have a significant shifting in $R_{max}$, contributing to the deviation of actual $F_{max}$ from the $F_{max}$ predicted by their $C_{v,0}$.
  (a) Initial wind profile of the Cyclone Amphan (2020) seed. 
  The blue curve represents $U_t$ profile, and the brown dashed curve depicts $U_{Rh}$ profile at the corresponding latitude. 
  Red dots denote the detected local maxima of $U_t$.
  (b) Hovmöller diagram displaying the temporal evolution of $U_t$ for the Cyclone Amphan (2020) seed. 
  The contours represent $U_t$, while the thick black curve illustrates the progression of $R_{max}$.
  (c) Same as (a) but for the seed of Cyclone Gati (2020).
  (d) Same as (b) but for the seed of Cyclone Gati (2020).}
  \label{fig:Outlier_Seed}
\end{figure*}

Reasons why vortices with similar $C_{v,0}$ values but different structures display notable differences in their weakening rates ($F_{max}$ spread) are not completely understood.
One scenario that can potentially result in deviations of members' $F_{max}$ from the theoretical prediction curve is the inward shifting of $R_{max}$.
This shift in $R_{max}$ may arise due to the presence of multiple local maxima of $U_t$ for the seed, or due to the seed's initial incompact structure.
Fig.\ref{fig:Outlier_Seed}a illustrates the radial wind profile of a member in the $ASYMSEED$, which initially exhibits two detected local maxima in $U_t$.
As the inner local maximum of $U_t$ yields a higher $C_v$ (with similar $U_t$ but at a smaller $R_{max}$), it is utilized to calculate $C_{v,0}$ for this seed.
However, since the outer local maximum of $U_t$ initially has a higher value, it is considered as the $V_{max}$ of this seed.
During the first $5$ days of the simulation, $R_{max}$ rapidly shifts inward as the outer circulation, including the outer local maximum of $U_t$, immediately weakens due to planetary Rossby wave drag (Fig.\ref{fig:Outlier_Seed}b).
Consequently, the predicted $F_{max}$ ($-4.9 \%/day$) based on $C_{v,0}$ overestimates (resulting in a less negative value) the actual $F_{max}$ ($-6.9 \%/day$) of this seed. 

On the other hand, when a seed is initially incompact, as demonstrated in Figure \ref{fig:Outlier_Seed}c, the initial $V_{max}$ of the seed experiences rapid weakening due to the impact of planetary Rossby wave drag, as illustrated in Fig.\ref{fig:Outlier_Seed}d.
Meanwhile, the inner circulation within the dynamical pouch experiences relatively less planetary Rossby wave drag. 
As a result, the inner circulation weakens less and thus becomes stronger than the outer $V_{max}$. 
This leads to the replacement of the weakening outer $V_{max}$ by the more persistent inner circulation as the new $V_{max}$ throughout the remainder of the simulation. 
This transition is clearly depicted by the abrupt jump in $R_{max}$ during the initial $2$ days of the simulation, as shown in Fig.\ref{fig:Outlier_Seed}d.
Since the calculation of $F_{max}$ is based on the evolution of $\tilde{V}_{max}$ during the first $5$ days, it incorporates both the rapid weakening of the initially outer $V_{max}$ and the more gradual weakening of the persistent inner $V_{max}$. 
Consequently, the predicted $F_{max}$ ($-12.1 \%/day$) based on $C_{v,0}$ underestimates (resulting in a more negative value) the actual $F_{max}$ ($-4.4 \%/day$) of this seed.

We are in pursuit of a straightforward theoretical explanation for the tight dependence of $F_{max}$ on $-C_{v,0}^{-1/2}$ found in our simulations.
Considering that all our members' weakening is governed by planetary Rossby wave drag, an equation for wave drag may offer a path towards a theory.
Here, we employ the horizontal component of the azimuthal mean tangential momentum budget equation on a barotropic $\beta$-plane in a storm-following cylindrical coordinate \citep{Kurihara_and_Tuleya_1974, Persing_et_al_2002}. 
In a quiescent environment, the equation is expressed as follows:
\begin{equation}\label{eq:Ut_budget}
\frac{\partial\overline{U_t}}{\partial t}=-\overline{U_r'\zeta'}-\beta r \overline{sin\theta(U_r'+c_r)},
\end{equation}
where $U_r$ is the radial wind, $\zeta$ is the vertical relative vorticity, $r$ is the radius, $\theta$ is the counterclockwise azimuth angle measured from the eastward direction, and $c_r$ is the radial component of the storm motion vector. Overbars indicate azimuthally averaged quantities, and primed quantities indicate deviations from this azimuthal average.
Note that the mean vorticity flux ($-\overline{U_r}\overline{\zeta}$) is neglected since there is no azimuthal mean divergent flow in a barotropic fluid.

When applied to vortex intensity, we can replace $U_t$ and $r$ in Eq.\ref{eq:Ut_budget} with $V_{max}$ and $R_{max}$, respectively.
Since our emphasis is on the role of planetary Rossby wave drag, we neglect the radial eddy vorticity flux at $R_{max}$ in Eq.\ref{eq:Ut_budget}.
Therefore, Eq.\ref{eq:Ut_budget} can now be simplified as follows:
\begin{equation}\label{eq:Vmax_budget}
\frac{\partial V_{max}}{\partial t}=-\beta R_{max} \overline{sin\theta(U_r'+c_r)}.
\end{equation}
By using the definition of $C_v$ (Eq.\ref{eq:Compactness}), we can express $R_{max}$ as follows:
\begin{equation}\label{eq:Rmax_Cv}
R_{max}=\sqrt{\frac{V_{max}}{2\pi\beta C_v}}.
\end{equation}
Substituting $R_{max}$ in Eq.\ref{eq:Vmax_budget} with Eq.\ref{eq:Rmax_Cv}, we obtain:
\begin{equation}\label{eq:Vmax_Cv}
\frac{\partial V_{max}}{\partial t}=-\sqrt{\frac{\beta V_{max}}{2\pi C_v}} \overline{sin\theta(U_r'+c_r)}.
\end{equation}
For the tendency of $V_{max}$ at the initial time step, Eq.\ref{eq:Vmax_Cv} can be written as:
\begin{equation}\label{eq:Vmax_Cv_ini}
\frac{\partial V_{max}}{\partial t}=-\sqrt{\frac{\beta V_{max,0}}{2\pi C_{v,0}}} \overline{sin\theta(U_r'+c_r)}.
\end{equation}
Thus, we can derive the tendency of $\tilde{V}_{max}$ by dividing both side of Eq.\ref{eq:Vmax_Cv_ini} with $V_{max,0}$, yielding:
\begin{equation}\label{eq:Nor_Vmax_Cv_ini}
\frac{\partial\tilde{V}_{max}}{\partial t}=-\sqrt{\frac{\beta} {2\pi V_{max,0} C_{v,0}}} \overline{sin\theta(U_r'+c_r)}.
\end{equation}
Subsequently, considering that the $LHS$ of Eq.\ref{eq:Nor_Vmax_Cv_ini} can be reasonably approximated as the average decreasing rate of $\tilde{V}_{max}$ over the initial $5$ days, we assume that $F_{max}$ is proportional to the initial tendency of $\tilde{V}_{max}$:
\begin{equation}\label{eq:Fit_F_Cv_long}
F_{max}\propto\frac{\partial\tilde{V}_{max}}{\partial t}\\
=-\sqrt{\frac{\beta} {2\pi V_{max,0} C_{v,0}}} \overline{sin\theta(U_r'+c_r)}\\
\propto-\sqrt{\frac{1}{C_v,0}}.
\end{equation}
This derivation yields the $-C_{v,0}^{-1/2}$ dependence seen in our experimental results above.
However, both $\beta$ and $V_{max,0}$ are also included Eq.\ref{eq:Fit_F_Cv_long}. 
While $\beta$ changes only minimally in our experiments, $V_{max,0}$ spans a wide range, which indicates that our prediction that only considers $C_{v,0}$ is an incomplete one.
Meanwhile, variations in $\overline{\sin\theta(U_r'+c_r)}$, which represents azimuthal variations in the eddy radial flow, hold potential significance and might offset the influence of $V_{max,0}$ term. 
As this eddy radial flow term lacks its own analytical framework, its evaluation within this context is not feasible and is instead characterized by the parameter $\alpha$ in Eq.\ref{eq:Fit_F_Cv}.
Ultimately, the fitted prediction curve explains the majority of variance in Fig.\ref{fig:Scatter}, which means that $F_{max}$ is dominant by $-C_{v,0}^{1/2}$, while $\beta$ and $V_{max,0}$ may induce additional secondary effects on the intensity evolution. 

\section{Conclusions and Discussions}
\label{Conclusions}

This study aims to investigate the influence of planetary Rossby wave drag on TC seed vortex intensity and its sensitivity to vortex structure. The following key findings have been established:
\begin{itemize}
    \item Vortex structural compactness ($C_v$) is defined as the ratio between vortex intensity ($V_{max}$) and the Rhines speed ($U_{Rh}$) at the radius of maximum wind ($R_{max}$), serving as an indicator of planetary Rossby wave drag's impact on $V_{max}$.
    \item Considering planetary Rossby wave drag as the sole vorticity sink, the initial $C_v$ ($C_{v,0}$) of a vortex exhibits a direct correlation with the strength of this wave drag on $V_{max}$. 
    Hence, $C_{v,0}$ has been established as a reliable predictor of the vortex weakening rate ($F_{max}$) induced by planetary Rossby wave drag.
    \item The analysis of $VARYRVMAX$ and $RANKINE$ demonstrates the accurate prediction of $F_{max}$ in idealized axisymmetric vortices on a barotropic $\beta$-plane based on its initial $C_{v,0}$ .
    \item The experiment sets $ASYMSEED$ and $SYMSEED$ further validate the applicability of our structural compactness theory to realistic vortices.
    \item In $ASYMSEED$, the inclusion of realistic 2-D asymmetry does not significantly alter the relationship between $C_{v,0}$ and $F_{max}$.
    \item $SYMSEED$ removes the 2-D asymmetry from the realistic seed vortex and produces similar results to $ASYMSEED$, but with a more robust $C_{v,0}$-$F_{max}$ relationship.
    \item In $LARGESEED$, we directly manipulate vortex $C_{v,0}$ by manually enlarging the entire vortex while keeping all other structural features unchanged.
    The results demonstrate that increasing the sample size through vortex enlargement yields a similar relationship between $C_{v,0}$ and $F_{max}$.
    \item The weakening rate in our experiments closely follows a $C_{v,0}^{-1/2}$ dependence, and is particularly robust when the vortex is initially compact ($C_{v,0}>1$).
\end{itemize}

In conclusion, the comprehensive experiment sets conducted on the barotropic $\beta$-plane provide robust evidence supporting the proposition that vortex structural compactness serves as a reliable predictor of the rate at which the vortex weakens due to planetary Rossby wave drag.
A TC seed vortex with a compact structure (higher $C_v$) experiences less planetary Rossby wave drag on their $V_{max}$, and consequently, they have the potential to persist longer compared to those with an incompact structure (lower $C_v$) under similar environmental conditions.

Across all of our experiment sets, even when a vortex possesses a $C_v$ value greater than $1$, it still experiences weakening due to planetary Rossby wave drag.
This observation underscores the fact that the dynamical pouch does not function as an impervious barrier that entirely shields $V_{max}$ from stimulating planetary Rossby wave.
In fact, planetary Rossby waves can be stimulated even when the circulation is within the dynamical pouch.
However, due to the relatively short circulation timescale within the dynamical pouch, the vortex's circulation is able to self-advect before being significantly affected by planetary Rossby wave drag, resulting in less weakening on the circulation.
Consequently, if a vortex exhibits a $C_v$ value that is not significantly larger than $1$, self-advection at $R_{max}$ is comparable in magnitude to planetary Rossby wave drag, albeit slightly greater.
As a result, the vortex will weaken due to planetary Rossby wave drag, while the weakening rate is much slower than those incompact vortices ($C_v<1$).

The parameter $C_v$ proposed in this study has important practical applications, particularly in the context of TC seed dynamics on Earth.
It plays a crucial role in determining a seed vortex's ability to resist significant drag by planetary Rossby waves, especially in the tropics. 
The requirement to maintain compactness in the seed vortex imposes limitations on its representation in weather and climate modeling.
For instance, when employing a global model with coarser horizontal resolution, the smaller values of $R_{max}$ and stronger magnitudes of $V_{max}$ associated with a compact seed vortex may not be adequately resolved.
Consequently, the model might underestimate the seed's $C_v$, leading to an overestimation of planetary Rossby wave drag and predicting a faster weakening for the seed.
This may explain why lower resolution models tend to produce far fewer tropical cyclones than expected \citep{Yamada_et_al_2021, Sobel_2021, Roberts_et_al_2020, Murakami_and_Sugi_2010}.
Ensuring accurate representation of compactness becomes essential in such modeling scenarios to obtain reliable predictions for TC seed evolution.

While we utilize a realistic TC seed structure obtained from ERA5 to initialize our barotropic model and examine the rate of vortex weakening, it is important to note that our results do not directly represent or imply the actual $V_{max}$ evolution of the seed in reanalysis data. 
The parameter $C_v$, as demonstrated in the preceding section, solely captures the influence of planetary Rossby wave drag on $V_{max}$, disregarding other physical processes, particularly the vorticity source term due to vortex stretching from convection \citep{Hsieh_et_al_2020}, that may potentially impact vortex intensity. 
Future work seeking to understand TC seeds in the real world should examine the competing effects of both source and sink terms of vorticity.

Several aspects of this subject remain unresolved, warranting further investigation.
While our theory and experimental results indicate the importance of $C_v$ as a parameter for examining the evolution of vortex intensity, the factors governing the evolution of vortex structural compactness remain uncertain.
The underlying causes behind the comparatively less robust prediction of $F_{max}$ for incompact vortices remain uncertain. 
Further exploration is needed to delve into the potential influence of other variations in vortex structure on this phenomenon.
Additionally, it remains uncertain whether developing seeds, destined for TC genesis, statistically possess a more compact structure compared to other seeds.
Addressing these inquiries necessitates modeling the development of TC seeds and conducting a comprehensive survey of TC seeds to better understand the interplay between seed structure and intensity.
By investigating these aspects, we can gain deeper insights into the dynamics and behavior of TC seeds, which may lead to advancements in our understanding of TC genesis and frequency.

\acknowledgments
The authors thank Tim Cronin for introducing us to the concept of wave drag. Thank James Penn and Geoffrey K. Vallis for posting their barotropic model code publicly. Funding support was provided by NSF Grant 1945113.
\datastatement
The data will be available on the Purdue University Research Repository (PURR). (IN PREP)

%






%



\bibliographystyle{ametsocV6}
\bibliography{Paper_Itself_V2}

\end{document}